\documentclass[%
 aip,
 amsmath,amssymb,
 reprint,%
]{revtex4-2}

\usepackage{graphicx}
\usepackage{dcolumn}
\usepackage{bm}
\usepackage[hidelinks]{hyperref}
\usepackage{bookmark}
\usepackage[utf8]{inputenc}
\usepackage[T1]{fontenc}
\usepackage{mathptmx}
\usepackage{bbm}
\usepackage{etoolbox}
\usepackage{braket}
\usepackage{siunitx}
 \DeclareSIUnit\angstrom{\text {Å}}
\usepackage{amsfonts}
\usepackage{amssymb}
\usepackage{upgreek}
\makeatletter
\def\@email#1#2{%
 \endgroup
 \patchcmd{\titleblock@produce}
  {\frontmatter@RRAPformat}
  {\frontmatter@RRAPformat{\produce@RRAP{*#1\href{mailto:#2}{#2}}}\frontmatter@RRAPformat}
  {}{}
}%
\makeatother

\newcommand{\lgt}{\text{Laguerre-Gaussian\ }}

\newcommand{\zray}{z_R}

\newcommand{\Udipole}{U_{\textsc{d}}}
\newcommand{\UZeeman}{U_{\textsc{z}}}

\renewcommand{\vec}{\boldsymbol}
\newcommand{\tf}{\textsc{\tiny{TF}}}
\newcommand{\rmi}{\textrm{i}}
\newcommand{\rme}{\textrm{e}}
\newcommand{\rmd}{\textrm{d}}
\newcommand{\Str}{S}
\newcommand{\br}{\vec{r}}
\newcommand{\mat}{\mathsf}

\newcommand{\rb}{\textsc{\tiny{Rb}}}
\newcommand{\ka}{\textsc{\tiny{K}}}

\newcommand{\Tr}[1]{\text{Tr}[#1]}

\usepackage{tikz}
\usetikzlibrary{shapes}
\DeclareRobustCommand\full  {\tikz[baseline=-0.6ex]\draw[thick] (0,0)--(0.5,0);}
\DeclareRobustCommand\dotted{\tikz[baseline=-0.6ex]\draw[thick,dotted] (0,0)--(0.54,0);}
\DeclareRobustCommand\dashed{\tikz[baseline=-0.6ex]\draw[thick,dashed] (0,0)--(0.54,0);}

\begin{document}

\preprint{AIP/123-QED}

\title[Efficient multipole representation for matter-wave optics]{Efficient multipole representation for matter-wave optics}
\author{J. Teske}
\email{jan.teske@physik.tu-darmstadt.de}
 \affiliation{Technical University of Darmstadt, Institute of Applied Physics, Darmstadt, Germany}
\author{R. Walser}%

\date{\today}

\begin{abstract}
Technical optics with matter waves requires a universal description of three-dimensional traps, lenses, and complex matter-wave fields. In analogy to the two-dimensional Zernike expansion in beam optics, we present a three-dimensional multipole expansion for Bose-condensed matter waves and optical devices. We characterize real magnetic chip traps, optical dipole traps, and the complex matter-wave field in terms of spherical harmonics and radial Stringari polynomials. We illustrate this procedure for typical harmonic model potentials as well as real magnetic and optical dipole traps. 
Eventually, we use the multipole expansion to characterize the aberrations of a ballistically interacting expanding Bose-Einstein condensate in (3+1)-dimensions. In particular, we find deviations from the quadratic phase ansatz in the popular scaling approximation. This universal multipole description of aberrations can be used to optimize matter-wave optics setups, for example in matter-wave interferometers.
\end{abstract}

\maketitle

\section{Introduction}
In 1934, Frits Zernike introduced the orthogonal "Kreisfl\"achenpolynome" to describe the optical path difference between light waves and a spherical reference wavefront \cite{ZERNIKE_1934_BeugungstheorieSchneidenverfahrens}. 
Understanding the phase differences and minimizing the optical aberrations laid the base for the first phase-contrast microscope \cite{ZERNIKE_1935_PhasenkontrastverfahrenBei}. This invention was awarded with the Nobel Prize in Physics in 1953. Nowadays, Zernike polynomials are widely used in optical system design as a standard description of imperfections in optical imaging \cite{GROSS_2005_HandbookOptical_VolI}. Typical wavefront errors are known as defocus, astigmatism, coma, spherical aberration, etc. \cite{GROSS_2006_HandbookOptical_Vol3}. Balancing aberrations is also relevant for optical imaging with electron microscopes \cite{HAIDER_1998_ElectronMicroscopy,BATSON_2002_SubangstromResolution, MULLER_2008_AtomicScaleChemical, ROSE_2009_GeometricalChargedParticle}. In contrast to visible light, massive particles, such as electrons, atoms, and even larger molecules \cite{CRONIN_2009_OpticsInterferometry}, have a much smaller de Broglie wavelength, $\lambda_{\text{dB}}=h/p$, and therefore a higher resolution.

Nowadays, atom-interferometers with ultracold atoms are used to study fundamental scientific questions like tests of the Einstein equivalence principle \cite{Ertmer2009,SCHLIPPERT_2014_QuantumTest, GABEL_2019_BoseEinsteinCondensates}, probing the quantum superposition on macroscopic scales \cite{KOVACHY_2015_QuantumSuperposition}, the search for dark matter candidates \cite{EL-NEAJ_2020_AEDGEAtomic} and gravitational waves \cite{DIMOPOULOS_2009_GravitationalWave,TINO_2019_SAGEProposal}. Being very sensitive to accelerations and rotations, atom interferometry could be used for inertial sensing, replacing commercial laser gyroscopes, and satellite navigation in space \cite{BONGS_2019_TakingAtom}. Common to all $\mu$g-interferometric measurements with Bose-Einstein condensates are long expansion times\cite{NANDI_2007_DroppingCold} to reduce mean-field interaction as well as to increase the sensitivity of the interferometer. Hence, it is crucial to understand the actual shape of the condensate's phase as it determines the interference patterns at the end of the interferometer \cite{NEUMANN_2021_AberrationsDimensional,NEUMANN_2021_AberrationsAtomic,CORNELIUS_2022_AtomInterferometry}. 

Inspired by Zernike's work, we will adopt his approach to analyze these aberrations in the world of matter waves:
 
First, we introduce a multipole expansion with suitable polynomial basis functions in 
Sec.~\ref{sec:Multipole expansion with Stringari polynomials}. 
We consider {{spherical-~,}} spheroidal-, 
displaced asymmetric harmonic- and generally asymmetric trapping potentials in Sec.~\ref{sec:Multipole expansion of trapping potentials}. 
In particular, we characterize the magnetic potential from a realistic atom chip model. In Sec.~\ref{sec:Multipole expansion of Bose-Einstein condensates}, we extend the multipole analysis to Bose-Einstein condensates in the strongly interacting Thomas-Fermi as well as in the low interacting limit. Finally, we investigate the shape of the phase profile for a ballistically expanding condensate in Sec.~\ref{sec:Expansion from magnetic chip traps} and conclude the discussion in Sec.~\ref{sec:Conclusion and outlook}.
\section{Multipole expansion with Stringari polynomials}
\label{sec:Multipole expansion with Stringari polynomials}
\subsection{Orthogonal function within a sphere}
Cold atoms can be trapped or guided in either optical dipole or Zeeman potentials \cite{METCALF_1999_LaserCooling}. If these potentials are applied for short times (impact approximation, phase imprinting \cite{DOBREK_1999_OpticalGeneration}), one modifies only the phase of the atomic wave packet, thus, the momentum distribution of the condensate. This process is equivalent to a \emph{thin lens} in optics, but now in (3+1) dimensions\cite{AMMANN_1997_DeltaKick}. Systematically analyzing the features of different potentials becomes crucial for achieving the ultimate precision in long-time atom interferometry \cite{ABEND_2023_TechnologyRoadmap}.

The standard Taylor expansion of a potential, in the vicinity of a point $\br_0$, in three-dimensional Cartesian coordinates $\br=\br_0+\vec{\zeta}$ reads 
\begin{align}
    U(\br) = & U_0 + \vec{\zeta}^\top 
    \nabla U_{|\br_0}  + \tfrac{1}{2}\vec{\zeta}^\top \mat K \vec{\zeta} + \ldots \label{eq:Taylor polynomials}.
\end{align}
This interpolation polynomial is useful for extracting forces from the gradient or trapping frequencies from the eigenvalues of the Hesse matrix $\mat K =(\nabla\otimes\nabla U)_{|\br_0}$. However, the Taylor series are notoriously inefficient approximation schemes. 

Alternatively, we introduce a multipole expansion in spherical coordinates $(r,\vartheta,\varphi)$
\begin{align}
	\label{eq:Multipole expansion}
	\braket{\br|U} = U(\br) =   & \sum_{n=0}^\infty\sum_{l=0}^\infty\sum_{m=-l}^l U_{nlm} \braket{\br|nlm},\end{align}
in terms of multipole coefficients $U_{nlm} = \braket{nlm|U}$ and orthonormal basis functions 
\begin{align}
	\label{eq:ortho Stringari}
    \braket{\br|nlm} &= \Str_{nlm}(\br) = \Str_{nl}(r) Y_{lm}(\vartheta,\varphi), \\
    \braket{n'l'm'|nlm} &=\delta_{n,n'} \delta_{l,l'}\delta_{m,m'},
    \end{align}
consisting of spherical harmonics $Y_{lm}(\vartheta,\varphi)$ and radial polynomials $\Str_{nl}(r)$ (see App.~\ref{sec:Jacobi polynomials}). Explicitly, they have support within a hard three-dimensional "aperture" of radius $R$. In terms of a scaled aperture radius $0\leq\tilde{r}=r/R\leq1$  , they read
\begin{equation}
	\label{eq:Stringaris}
    S_{nl}(r) =\mathcal{N}_{nl}\,  \tilde{r}^{l} J_n^{(l+1/2, 0)}(1-2\tilde{r}^2).
\end{equation}
The polynomials in Eq.~\eqref{eq:Stringaris} describe the sound wave excitations of a Bose-Einstein condensate in the strong interacting 
Thomas-Fermi limit for an isotropic three-dimensional harmonic oscillator potential, originally described by S.~Stringari\cite{STRINGARI_1996_CollectiveExcitations} and P.~Öhberg et al.\cite{P.OHBERG_1997_LowenergyElementary}. Thus, these fundamental modes are well adapted for condensates in harmonic as well as in more realistic trapping geometries (see Sec.~\ref{sec:Magnetic Zeeman potential of an atom chip} and~\ref{sec:Expansion from magnetic chip traps}).
In hindsight, it almost seems natural to recover Jacobi polynomials $J_n^{(l+1/2, 0)}$ with half-integral indices in the three-dimensional situation, if we compare this to the  circle polynomials of F.~Zernike in two-dimensional beam optics \cite{BORN_1999_PrinciplesOptics}. 
The normalization $\mathcal{N}_{nl}  = [(4n + 2l + 3)/R^{3}]^{1/2}$ renders the polynomials orthonormal on the interval $0\leq r \leq R$,
\begin{equation}\label{eq:Ortho Stringari}
 \int_0^R \rmd r~  r^2 S_{n'l}(r) S_{nl}(r)  = \delta_{n,n'}. 
\end{equation}
Therefore, the potential expansion coefficients $U_{nlm}$ in Eq.~\eqref{eq:Multipole expansion} are given by the overlap integral
\begin{equation}
\label{eq:Multipole coefficients by projection}
    U_{nlm} = \braket{nlm|U}= \int_{V} \rmd^3 r~ \Str^*_{nlm}(\vec r) U(\br),
\end{equation}
within a spherical volume $V=4\pi R^3/3$. Typically, the characteristic size of the wave packet in the trap determines the aperture radius $R$. For cold clouds, this size is of the order of the harmonic oscillator length $\ell=\sqrt{\hbar/(M\omega)}$ for a non-interacting gas or of the Thomas-Fermi radius $r_\tf=\sqrt{2\mu /(M\omega^2)}$ for an interacting condensate. For thermal clouds, the radius $R\approx \sqrt{k_B T/(M \omega^2)}$ is proportional to the temperature $T$.

Instead of a multipole expansion of the potential in Eq.~\eqref{eq:Dipole potential exponential form}, it is also possible to decompose the logarithm of the potential
\begin{equation}
	\label{eq:multipole expansion cumulant}
    \theta(\br)=-\log{\frac{U(\br)}{U_0}}=\sum_{n=0}^\infty\sum_{l=0}^\infty\sum_{m=-l}^l \theta_{nlm} \braket{\br|nlm}.
\end{equation}
This cumulant expansion \cite{GARDINER_1985_HandbookStochastic} is particularly useful for Gaussian functions as the series terminates quickly after the second order. In Sec.~\ref{sec:Multipole expansion of trapping potentials}, we analyze both the potentials and their cumulants in terms of these multipole expansions.
\subsection{Spectral powers}
For interpreting and visualizing the complex expansion coefficient  $U_{nlm}$ or $\theta_{nlm}$, we introduce relative spectral powers, $p_{nl}\geq 0$. The latter is independent of the potential's orientation to the reference coordinate system. This also holds for the cumulant expansion in Eq.~\eqref{eq:multipole expansion cumulant}. Let us consider a coordinate transformation $\br' = \mat{R} \br$, where $\mat R$ denotes an orthogonal rotation matrix $\mat{R}\mat{R}^\top=\mathbbm{1}$. In both frames, the values of the potential agree
\begin{equation}
    \braket{\br|U} = \braket{\br'|U'}=\sum_{nlm'} \braket{\br'|nlm'} U'_{nlm'}.
\end{equation}
The coefficients in the new reference frame $U'_{nlm}$ are given by
\begin{equation}
    U'_{nlm'} =\sum_{m=-l}^l  D^{(l)}_{m'm}(\mat R) U_{nlm},
\end{equation}
where the $D^{(l)}_{m'm}(\mat R)=\braket{lm'|\hat{R}|lm}$ are the Wigner D-matrices as the matrix representation of the rotation operator $\hat{R}$ in the angular momentum basis \cite{VARSHALOVICH_1988_QuantumTheory,BALLENTINE_2014_QuantumMechanics,GALINDO_1990_QuantumMechanics}. Using the unitarity of the rotation operator, $\hat{R}\hat{R}^\dag=\hat{R}^\dag\hat{R}=\mathbbm{1}$, as well the orthonormality of the Stringari polynomials \eqref{eq:ortho Stringari}, we specify rotational invariant measures such as the total power
\begin{equation}\label{eq:Total power}
	P(U)  =\int_{V} \rmd^3 r~ |U(\br)|^2=\sum_{nlm}|U_{nlm}|^2.
\end{equation}
Moreover, we define the marginals $P_{nl}$, $P_{l}$, 
\begin{align}
    P_{nl}(U)&=\sum_{m=-l}^l|U_{nlm}|^2, & P_l(U)&=\sum_n P_{nl}(U),\\
    \intertext{as well as the relative fractional powers $p_{nl}(U)$ by}
    p_{nl}(U)&=\frac{P_{nl}(U)}{P(U)} &  \sum_{nl}p_{nl}(U)=& 1. 
\end{align}
Thus, the  relative fractional powers add up to one. 
\section{Multipole expansion of traps}
\label{sec:Multipole expansion of trapping potentials}
\subsection{Harmonic, isotropic, three-dimensional oscillator}
\label{sec:Harmonic, isotropic, three-dimensional oscillator}
A stiffness parameter $k$ characterizes a harmonic,  isotropic three-dimensional oscillator potential 
\begin{align}
    \label{eq:Spherical harmonic trap}
    U(\br) = 
    \frac{k}{2}
    r^2&= \sum_{n=0}^1 U_{n00} \braket{\br|n00} \nonumber \\
    &= \tfrac{3}{5}u\braket{\br|000} - \sqrt{\tfrac{12}{175}}u\braket{\br| 100}.
\end{align}
For a particle with mass $M$, the angular frequency $\omega = \sqrt{k/M}$.  An isotropic potential is invariant under rotations. Thus, only the s-waves contribute. A Stringari polynomial $\Str_{n0}$ from Eq.~\eqref{eq:Stringaris} is of the order $r^{2n}$, which limits the radial modes to two monopoles. The expansion coefficients $U_{n00}$ are given in terms of a dimensional factor [\si{\joule\meter\tothe{3/2}}]
\begin{equation}
	u = \frac{k}{2} R^2 \sqrt{V}.
\end{equation}
For the total power, one finds 
\begin{align}
	\label{eq:total and fractional powers isotropic}
    P(U) =&  \frac{3}{7} u^2,
\end{align}
as well as the fractional powers
\begin{equation}
\label{eq:pisotropictrap}
    p_{00}(U) =\frac{21}{25}=0.84, \quad p_{10}(U)=\frac{4}{25}=0.16.
\end{equation}
\subsection{Harmonic, anisotropic, three-dimensional oscillator}
\label{sec:Harmonic, anisotropic, three-dimensional oscillator}
\subsubsection{Spheroidal potential}
\label{sec:Prolate or oblate spheriodal potential}
For a prolate or oblate spheroidal harmonic oscillator, the potential is characterized by two 
stiffness constants $k_\perp$ and $k_\parallel$, but it is still rotational symmetric around the z-axis
\begin{equation}\label{eq:spheriodal harmonic trap}
    U(\br)=\frac{k_\perp}{2}(x^2+y^2+\alpha^2 z^2). 
\end{equation}
The anisotropy is measured by $\alpha^2=k_\parallel/k_\perp$. In contrast to the isotropic oscillator, one needs also a quadrupole within the multipole expansion
\begin{align}\label{eq:Multipoles prolate oblate}
    \begin{aligned}
     U _{000}&= \frac{2+\alpha^2}{5} u,&
    U_{100} &= -\frac{4+2\alpha^2}{\sqrt{525}}  u ,\\
    U_{020} &= -\frac{2-2\alpha^2}{\sqrt{105}} u,&
    u &=  \frac{k_\perp}{2} R^2 \sqrt{V}.   
    \end{aligned}
\end{align}
Using the coefficients in Eq.~\eqref{eq:Multipoles prolate oblate} we can evaluate the total power
\begin{equation}
    P(U)= \frac{8 + 4 \alpha^2 + 3 \alpha^4}{35} u^2,
\end{equation}
and the relative fractional powers
\begin{align}
    \begin{aligned}
     p_{00}(U) &= \frac{7}{5}\frac{(2+\alpha^2)^2}{8 + 4 \alpha^2 + 3 \alpha^4},  \\ 
     p_{10}(U)&=\frac{4}{15}\frac{(2+\alpha^2)^2}{8 + 4 \alpha^2 + 3 \alpha^4}, \\
     p_{02}(U)&=   \frac{4}{3} \frac{1-\alpha^2}{8 + 4 \alpha^2 + 3 \alpha^4},
    \end{aligned}
\end{align}
for the cylindrical symmetric oscillator potential. It is noteworthy that this expansion encompasses degenerate traps with vanishing z-confinement $\alpha = 0$, as well as isotropic traps with $\alpha = 1$ from Eq.~\eqref{eq:pisotropictrap}.
\begin{figure*}
    \centering
    \includegraphics{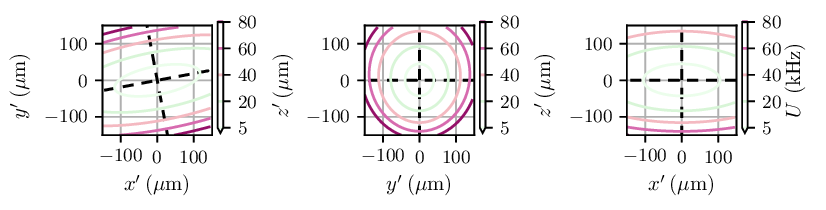}
    \caption{Magnetic Zeeman potential of an atomic chip trap for $^{87}$Rb in the magnetic substate $\ket{F=2,m_F=2}$. Two-dimensional contour plots along all three spatial planes at position $\br' =\br - \br_0$. The principle axis of the trap obtained by the Hesse matrix $\mat{K}$ Eq.~\eqref{eq:Taylor polynomials} marked as ~ \dashed ~ lines. Parameters of the trap are summarized in Tab.~\ref{tab:release trap currents} and~\ref{tab:release trap parameters}.}
    \label{fig:Isolines Zeeman potential release trap}
\end{figure*}
\begin{figure*}
    \centering
    \includegraphics[scale=1]{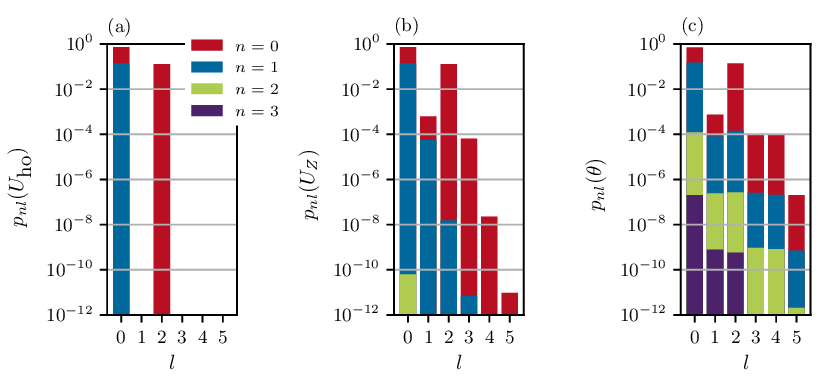}
    \caption{Multipole expansion of the magnetic chip trap potential. Relative powers $p_{nl}(U)$ versus angular momentum $l$ of the Zeeman potential shown in Fig.~\ref{fig:Isolines Zeeman potential release trap}. Different principle numbers: red $n=0$, blue $n=1$, green $n=2$, purple $n=3$. (a) harmonic approximation $p_{nl}(U_{\textrm{ho}})$, (b) Zeeman potential $p_{nl}(\UZeeman)$, (c) cumulant $p_{nl}(\theta)$. We used $R=\SI{40}{\micro \meter}$ and $n_{\text{max}}=3$, $l_{\text{max}}=5$ in Eq.~\eqref{eq:Stringaris},~\eqref{eq:Multipole expansion}, and~\eqref{eq:multipole expansion cumulant}.}
    \label{fig:relative powers magnetic Zeeman potential}
\end{figure*}
\subsubsection{Tilted, shifted anisotropic  harmonic oscillator potential}
We consider a general harmonic oscillator potential  with a symmetric stiffness matrix $\mat{K}$ localized at a position  $\vec{r}_0$
\begin{align}
    U(\br,\br_0)=&(\br-\vec{r}_0)^\top \tfrac{\mat{K}}{2} (\br-\vec{r}_0) =
    \vec{r}_0^\top \tfrac{\mat{K}}{2} \vec{r}_0 -\vec{r}_0^\top \mat{K} \vec{r}+  \br^\top \tfrac{\mat{K}}{2} \br \notag \\
      =& U^{(0)}(\br_0) + U^{(1)}(\br,\br_0) + U^{(2)}(\br), 
      \label{eq:multipole components harmonic shifted trap}
\end{align}
and gather it into homogeneous potentials  $U^{(n)}(\br)$ of degree $n$. 
In order to determine the multipole expansion, we transform the  position vectors $\br=x\vec{e}_x+y\vec{e}_y+z\vec{e}_z$ given in a Cartesian basis
$\{\vec{e}_{x},\vec{e}_{y}, \vec{e}_{z}\}$ to the spherical basis $\{\vec{e}_{1},\vec{e}_{0}, \vec{e}_{-1}\}$ 
\begin{align}
\label{eq:sphericalbasisvectors}
    \vec{e}_{1} &=-\frac{\vec{e}_x+ \rmi \vec{e}_y}{\sqrt{2}} , &
    \vec{e}_0&=\vec{e}_z, &
        \vec{e}_{-1} &=\frac{\vec{e}_x- \rmi \vec{e}_y}{\sqrt{2}},
\end{align}
which are orthogonal with respect to the standard complex scalar-product $\vec{e}^{}_n \, \vec{e}_m^*=\delta_{n,m}$. It is convenient to introduce also a dual basis $\{\vec{e}^{1},\vec{e}^{0}, \vec{e}^{-1}\}$ with
\begin{align}
\label{eq:sphericalbasisdual}
    \vec{e}^m&= (-1)^m\vec{e}_{-m}=\vec{e}_m^\ast ,\\
    \vec{e}_n \, \vec{e}^m &= \vec{e}^{}_n \, \vec{e}_m^*=\delta_{n,m}.
\end{align}
Now, the position vector reads in the co- and contravariant spherical basis  \cite{VARSHALOVICH_1988_QuantumTheory}
\begin{align}\label{eq:spherical basis}
    \br&=\sum_{m=-1}^1 q^m\vec{e}_m=\sum_{m=-1}^1 q_m \vec{e}^m,  \\
q_{\pm 1}&=\mp \frac{x\pm \rmi y}{\sqrt{2}} , \qquad q_0=z. 
\end{align}
More generally, one can express the spherical vector components \cite{EDMONDS_1957_AngularMomentum}  $q_m$  by 
\begin{equation}
    q_m = \left(\frac{4 \pi}{3}\right)^{1/2} Y_{1m}(\vartheta, \varphi) r = T^{(1)}_m(\vartheta, \varphi) r,
\end{equation}
with the spherical tensor $T^{(1)}_m(\vartheta, \varphi)$.
 
Obviously, the constant $U^{(0)}(\br_0)= \vec{r}_0^\top \mat{K} \vec{r}_0 /2$. The dipole coefficients for $U^{(1)}(\br,\br_0)$ in Eq.~\eqref{eq:multipole components harmonic shifted trap} are then given by the spherical components of the vector $\br_0$
\begin{align}
    U^{(1)}(\br,\br_0) &=  r \sum^{1}_{m=-1} U^{(1)}_{1m}(\br_0) T^{(1)}_{m}(\hat{r}), \\ U^{(1)}_{1m}(\br_0)&=-\sum_{s=-1}^1q^*_{0,s} \mat{K}_{ms}.
\end{align}
The second order contribution $U^{(2)}(\br)$ in Eq.~\eqref{eq:multipole components harmonic shifted trap} contains a product of two spherical tensors that can be simplified using the Clebsch-Gordan expansion \cite{EDMONDS_1957_AngularMomentum}
\begin{equation}\label{eq:Products spherical tensors}
    T_{m_1}^{(l_1)}(\hat{r}) T_{m_2}^{(l_2)}(\hat{r}) = \sum_{l=|l_1-l_2|}^{l_1+l_2}\sum_{m=-l}^l C^{l_1l_2l}_{m_1m_2m} T_{m}^{(l)}(\hat{r}),
\end{equation}
with Clebsch-Gordan coefficients\cite{GALINDO_1990_QuantumMechanics} $C^{l_1l_2l}_{m_1m_2m}=\braket{l_1m_1l_2m_2|lm}$. Hence, we can rewrite $U^{(2)}(\br)$ as
\begin{align}\label{eq:quadrupole general harmonic oscillator}
 U^{(2)}(\br) 
  &=\frac{r^2}{2} \sum_{l=0}^2\sum_{m=-2}^2 U^{(2)}_{lm}T_{m}^{(l)}(\hat{r}), \\
  U^{(2)}_{lm} &= \sum_{r,s=-1}^{1} (-1)^s 
    C^{11l}_{r(-s)m} \mat{K}_{rs},
\end{align}
with multipole coefficients that contain the matrix elements of $\mat K$ in the spherical basis and Clebsch-Gordan coefficients.
For the radial part, one can use the results from the expansion of isotropic oscillator \ref{sec:Harmonic, isotropic, three-dimensional oscillator} in terms of the s-wave Stringari polynomials. The total power in Eq.~\eqref{eq:Total power} of $U^{(2)}$ evaluates to
\begin{align}
    P(U^{(2)})&=9\frac{2 \Tr{\mat{K}^2} +\Tr{\mat K}^2}{35 \Tr{\mat{K}}^2}u^2,
    &
    u&= \frac{\Tr{\mat{K}}}{6} R^2 \sqrt{V}.
    \label{eq:total power anistropic} 
\end{align} 
For an isotropic stiffness $\mat{K}=k\mathbbm{1}$, Eq.~\eqref{eq:total power anistropic} reduces to the result of the isotropic harmonic oscillator in Eq.~\eqref{eq:total and fractional powers isotropic}. 
\subsection{Magnetic Zeeman potential of an atom chip}
\label{sec:Magnetic Zeeman potential of an atom chip}
Magnetizable atoms can be trapped in a static magnetic field \cite{BERGEMAN_1987_MagnetostaticTrapping}
\begin{equation}\label{eq:magPot}
    \UZeeman(\br)  =  \mu_B m_F  g_F |\vec{B}(\br)|.
\end{equation}
Here $\mu_B$ denotes the Bohr magneton, $g_F$ the Land\'e factor, $m_F$ the magnetic quantum number of the total angular momentum, and  $\vec{B}(\br)$ the magnetic induction field. In this work, we analyze the Zeeman potential in Eq.~\eqref{eq:magPot} obtained by a real-world model of a magnetic chip trap \cite{FORTAGH_2007_MagneticMicrotraps}. These have become popular when a miniaturization of the experimental setup is required, for example at drop towers \cite{MUNTINGA_2013_InterferometryBoseEinstein,DEPPNER_2021_CollectiveModeEnhanced} and in space \cite{LACHMANN_2021_UltracoldAtomb,FRYE_2021_BoseEinsteinCondensate}. The atom chip and its functionalities are described in detail in Ref.~\cite{RUDOLPH_2015_HighfluxBEC}. In App.~\ref{sec:Chip model}, we present a realistic finite-wire model of the microtrap from which we deduce the magnetic induction field (see Eq.~\eqref{eq:biot-savart-superposition}). Here, we consider $^{87}$Rb atoms in the magnetic hyperfine state $\ket{F=2,m_F=2}$ with $g_F=1/2$ with the set of currents as presented in Tab.~\ref{tab:release trap currents}. The corresponding Zeeman potential, evaluated in the chip coordinate system, is depicted in Fig.~\ref{fig:Isolines Zeeman potential release trap}.

For an efficient representation of the three-dimensional Zeeman potential, we use the multipole expansion in Eq.~\eqref{eq:Multipole expansion}. For a comparison, we extract also the multipoles of the cumulant as discussed in Eq.~\eqref{eq:multipole expansion cumulant}. As we represent the potential on discrete lattice points, we use a least-square optimization (see App.~\ref{sec:Numerical evaluation}) to calculate the expansion coefficients $U_{nlm}$ and $\theta_{nlm}$ respectively. The results of the multipole expansion are summarized in Fig.~\ref{fig:relative powers magnetic Zeeman potential}. There, we show the relative fractional angular powers $p_{nl}$ for the harmonic approximation $p_{nl}(U_{\textrm{ho}})$ (a), the full Zeeman potential $p_{nl}(\UZeeman)$ (b) and the cumulant $p_{nl}(\theta)$ (c). In each subfigure, we have used the same number of basis functions with maximal principle and angular momentum quantum numbers $n_{\text{max}}=3$, $l_{\text{max}}=5$. Moreover, the multipole expansion is performed at the position of the trap minimum, shifting the position vector in Eq.~\eqref{eq:multipole components harmonic shifted trap} by $\br =\br' + \br_0$. The latter implies a vanishing dipole component $U^{(1)}(\br,\br_0)=0$ for the harmonic approximation.

As discussed in Sec.~\ref{sec:Harmonic, anisotropic, three-dimensional oscillator}, the anisotropic harmonic oscillator potential exhibits just two monopoles and one quadrupole contributions, depicted in Fig.~\ref{fig:relative powers magnetic Zeeman potential} (a). 
For a real Z-wire trap on the atom chip, the Zeeman potential exhibits all multipoles: in particular the monopoles $p_{00}$, $p_{10}$,  dipoles with $p_{01}$, $p_{11}$, a quadrupole $p_{02}$, as well as the octupole with $p_{03}$, which is depicted in Fig.~\ref{fig:relative powers magnetic Zeeman potential} (b). From the dipole coefficients, we deduce that in the anharmonic trap, the position of the trap minimum does not coincide with the center-of-mass position of the trap. Thus, the application of a Zeeman lens-potential causes a finite momentum-kick to the atomic density distribution. While the dipoles affect the center of mass motion, the additional octupole causes density distortions in long-time matter-wave optics and interferometry. 

Finally, we note that multipoles of higher order $l>3$ are decreasing rapidly for our trap configuration. Comparing the direct multipole expansion to the cumulant expansion, one finds that the expansion of the cumulant series converges more slowly due to the logarithmic character of Eq.~\eqref{eq:multipole expansion cumulant}, see Fig.~\ref{fig:relative powers magnetic Zeeman potential} (c).
 
\subsection{Optical dipole potential from Laguerre-Gaussian beams}\label{sec:Optical dipole potential from lgt beams}
Besides magnetic trapping, atoms can be also trapped by an optical dipole potential
\cite{kazantsev90,Marksteiner1995}
\begin{align}
	\label{eq:Dipole potential exponential form}
	\Udipole(\br) &= U_0\, \rme^{-\theta(\br)},
	&
	U_0&=\hbar|\Omega_0|^2/(4 \Delta),
\end{align}
which is created by laser light far detuned from the atomic resonance. Here, $\Delta=\omega_L - \omega_0$ describes the laser detuning and $\Omega_0$ is the Rabi frequency. For red-detuned lasers $\omega_L<\omega_0$ with respect to the atomic transition $\omega_0$, the dipole potential is attractive and it is repulsive for blue-detuning $\omega_L>\omega_0$  \cite{GRIMM_2000_OpticalDipole}. 

\begin{figure}
    \centering
    \includegraphics[scale=0.90]{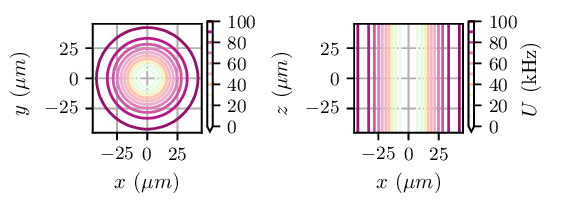}
    \caption{Optical dipole potential for a single Laguerre-Gaussian laser beam. Two-dimensional contour plots along all two spatial planes. Parameters of the trap as in reference \cite{KANTHAK_2021_TimedomainOptics}, trap depth  $|U_0|/k_B=\SI{5}{ \micro \kelvin} $,  Rayleigh range $z_R=\SI{3.2}{\milli \meter}$, $w_0=\SI{33}{\micro \meter}$, and the trapping frequencies $\vec{\nu}=(\num{211.0}, \num{211.0}, \num{1.5}) \si{\hertz} $ for $^{87}$Rb.} 
    \label{fig:Isolines single laser dipole potential}
\end{figure}

For a single \lgt beam \cite{MILONNI_2010_LaserPhysics} the exponent $\theta(\br)$ in Eq.~\eqref{eq:Dipole potential exponential form} has the spatial dependence 
\begin{equation}\label{eq:Single LG beam}
    \theta(\br) = 2\frac{x^2+y^2}{w(z)^2}
    +\ln{
        \left(1+\frac{z^2}{\zray^2}\right)}
    \approx 2\frac{x^2+y^2}{w_0^2}+\frac{z^2}{\zray^2},
\end{equation}
where the Rayleigh range $\zray= \pi w_0^2/\lambda_L,$ is typically much larger than the extension of the condensate wave packet. The laser wavelength is $\lambda_L=2\pi/k_L$  and the beam waist is denoted by $w(z)=w_0 (1+z^2/\zray^2)^{1/2}$. The dipole potential Eq.~\eqref{eq:Single LG beam} describes an optical waveguide \cite{BONGS_2001_WaveguideBoseEinstein} or can act as an optical matter-wave lens in the time domain \cite{KANTHAK_2021_TimedomainOptics}. For the latter, we depict the optical potential for $^{87}$Rb in Fig.~\ref{fig:Isolines single laser dipole potential}. 

The harmonic approximation of Eq.~\eqref{eq:Single LG beam}, corresponds to the spheroidal trapping potential in Eq.~\eqref{eq:spheriodal harmonic trap} with stiffness 
\begin{equation}
    k_\perp = \frac{4U_0}{w_0^2},\quad k_\parallel=\frac{2U_0}{\zray^2},
\end{equation}
and the anisotropy $\alpha^2 = w_0^2 /(2\zray)$ depending on the ratio of the minimal waist and the Rayleigh length.

\begin{figure*}
    \centering
    \includegraphics[scale=1]{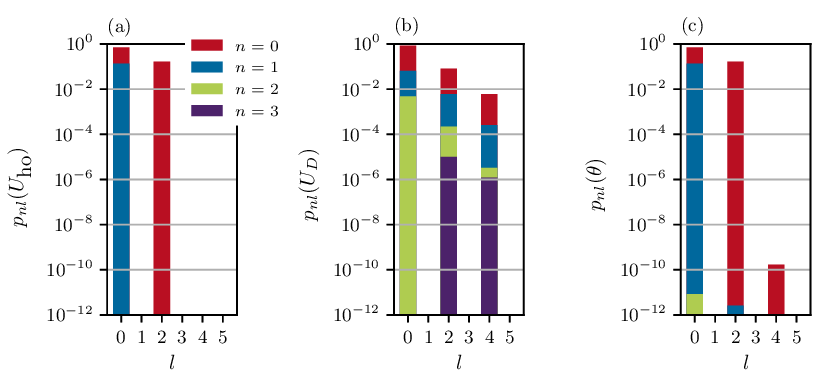}
    \caption{Multipole expansion of the optical dipole potential for a single Laguerre-Gaussian beam shown in Fig.~\ref{fig:Isolines single laser dipole potential}. Relative powers $p_{nl}(U)$ versus angular momentum $l$. Different principle numbers: red $n=0$, blue $n=1$, green $n=2$, purple $n=3$. (a) harmonic approximation $p_{nl}(U_{\textrm{ho}})$, (b) optical dipole potential $p_{nl}(\Udipole)$, (c) cumulant $p_{nl}(\theta)$. We used $R=\SI{40}{\micro \meter}$ and $n_{\text{max}}=3$, $l_{\text{max}}=5$ in Eq.~\eqref{eq:Stringaris},~\eqref{eq:Multipole expansion}, and~\eqref{eq:multipole expansion cumulant}.}
    \label{fig:relative powers single laser beam}
\end{figure*}

As in the previous subsection, we evaluate the relative powers, see Fig.~\ref{fig:relative powers single laser beam}, for the harmonic approximation $p_{nl}(U_{\textrm{ho}})$ Eq.~\eqref{eq:spheriodal harmonic trap} (a), the dipole potential $p_{nl}(\Udipole)$ Eq.~\eqref{eq:Dipole potential exponential form} (b) and the cumulant $p_{nl}(\theta)$ in Eq.~\eqref{eq:Single LG beam} (c). Due to the Gaussian laser beam, the cumulant expansion of the dipole potential is much more efficient than the direct multipole expansion (compare Fig.~\ref{fig:relative powers single laser beam} (b) and (c)). For large Rayleigh length $\zray \gg R$, the spatial dependence of the exponent $\theta(\br)$ is almost Gaussian, which is shown in subfigures~\ref{fig:relative powers single laser beam} (a) and (b).  Additional corrections to the harmonic cumulant in higher angular momentum components $l>2$ are of the order $10^{-9}$ and smaller. 
\section{Multipole expansion of Bose-Einstein condensates}
\label{sec:Multipole expansion of Bose-Einstein condensates}
Besides the external trapping potentials, we are also interested in an efficient representation of a three-dimensional Bose-Einstein condensate. Within the classical field approximation for Bosons, the evolution of the complex matter-wave field is described 
\begin{align}\label{eq:time-dependent GP equation}
\rmi \hbar \partial_t \Psi(\br,t) &= \left[-\frac{\hbar^2\vec{\nabla}^2}{2M} + U(\br,t) + g n(\br,t) \right] \Psi(\br,t), \\
N&=\int \rmd^3r ~ n(\br,t) ,
\end{align}
by the time-dependent Gross-Pitaevskii equation. Here, the contact-interaction strength is denoted by  $g=4 \pi \hbar^2 a_s/M$, with the atomic mass $M$ and the s-wave scattering length $a_s$. The atomic density $n(\br,t)=|\Psi(\br,t)|^2$  is normalized to the particle number $N$. Equivalently to Eq.~\eqref{eq:time-dependent GP equation}, one can represent the classical complex Gross-Pitaevskii field  in terms of the two real hydrodynamical variables, the density of the condensates $n(\vec{r},t)$ and its phase $\phi(\vec{r},t)$, 
\begin{equation}\label{eq:Madelung transform}
    \Psi(\br, t) = \sqrt{n(\br,t)}\rme^{\rmi \phi(\br,t)},
\end{equation}
also known as a Madelung transform \cite{DALFOVO_1999_TheoryBoseEinstein}.
 
To represent the three-dimensional matter-wave field, we choose the same multipole expansion as in Eq.~\eqref{eq:Multipole expansion},
\begin{align}
    n(\br,t) = \sum_{n=0}^{\infty}\sum_{l=0}^\infty\sum_{m=-l}^ln_{nlm}(t)\braket{\br|nlm},\label{eq:multipole expansion density} \\
    \phi(\br,t) = \sum_{n=0}^{\infty}\sum_{l=0}^\infty\sum_{m=-l}^l\phi_{nlm}(t)\braket{\br|nlm},\label{eq:multipole expansion phase}
\end{align}
for the density as well as for the phase.

In the stationary case, the time-dependent field is governed by $\Psi(\br,t) = \Psi(\br)\rme^{-\rmi \mu t /\hbar}$, where $\mu(N)$ is the chemical potential of the condensate. Thus, the Gross–Pitaevskii equation for stationary field $\Psi(\br)$ reads
\begin{equation}\label{eq:stationary GP}
\left[-\frac{\hbar^2\vec{\nabla}^2}{2M}  + U(\br) + g n(\br)\right]\Psi(\br) = \mu \Psi(\br).
\end{equation}
In the limit of large repulsive interactions, one can neglect the quantum pressure that arises from the localization energy of the classical field
\begin{equation}\label{eq:TF Field}
\left[U(\br) + gn_\tf(\br) - \mu_\tf \right]\Psi_\tf(\br) = 0.
\end{equation}
Eq.~\eqref{eq:TF Field} admits algebraic solutions of the form
\begin{equation}\label{eq:Thomas-Fermi density}
n_\tf(\vec r)=\left\lbrace
\begin{array}{rl}
\frac{\mu_\tf - U(\vec r)}{g}, & \mu_\tf - U(\vec r) \geq 0,\\ 
0 & \textrm{else},
\end{array}\right.
\end{equation}
known as the Thomas-Fermi approximation \cite{BAYM_1996_GroundStateProperties}.

In the following, we apply the multipole expansion in Eq.~\eqref{eq:multipole expansion density} for the condensate density in some of the trapping potentials discussed in the previous Sec.~\ref{sec:Multipole expansion of trapping potentials}. Thereby, we discuss the strongly interacting Thomas-Fermi regime as well as the exact numerical solution of the stationary Gross-Pitaevskii equation.

We should note that in principle the Stringari polynomials in Eq.~\eqref{eq:multipole expansion density} can exhibit negative values which are nonphysical when regarding positive-valued atomic densities. In order to avoid this anomaly, we consider coordinates in the interval $0 \leq r \leq R$ and densities $n(\br)\geq 0$. 

For the Thomas-Fermi density in Eq.~\eqref{eq:Thomas-Fermi density}, we expect an exact interpolation by the Stringari basis functions, if the potential is of polynomial form. In contrast, setting the interaction strength in Eq.~\eqref{eq:stationary GP} to $g=0$ and considering a harmonic oscillator potential, one obtains a Gaussian density distribution as the harmonic oscillator ground state \cite{GALINDO_1990_QuantumMechanics}
\begin{equation}\label{eq:density HO}
n(\br)=n_0\,\rme^{-\theta(\br)}, \quad \theta(\br)=\sum_{j=1}^3 \frac{r_j^2}{2\sigma_j^2}.
\end{equation}
The widths of the Gaussian correspond to the harmonic oscillator lengths $\sigma_j=\sqrt{\hbar/(M\omega_j)}$, $j=1,2,3$ for the three spatial directions. For the latter, the cumulant expansion 
\begin{equation} \label{eq:multipole expansion cumulant density}
    \theta(\br) = - \log \frac{n(\br)}{n_0}=\sum_{n=0}^\infty\sum_{l=0}^\infty\sum_{m=-l}^l \theta_{nlm} \braket{\br|nlm},
\end{equation}
is always of a quadratic form. Thus we recover the expansion coefficients for the optical dipole potential of a single Laguerre-Gaussian beam in Sec.~\ref{sec:Optical dipole potential from lgt beams}.
\subsection{Isotropic, three-dimensional density}
\begin{figure*}
    \centering
    \includegraphics{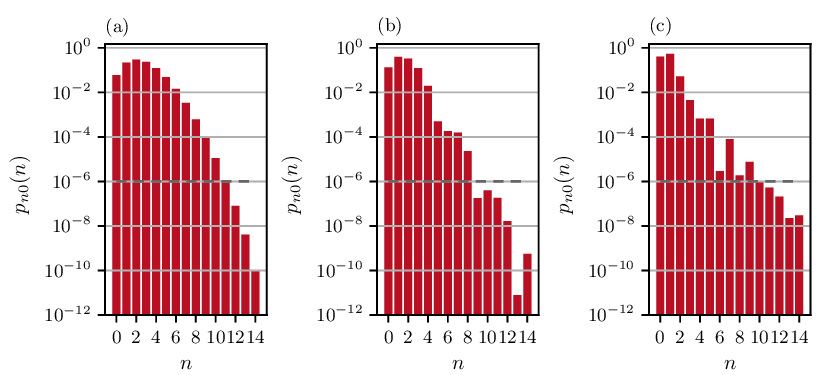}
    \caption{Multipole expansion of the Gross-Pitaevskii density $n(\br)$ Eq.~\eqref{eq:stationary GP} for the isotropic harmonic oscillator Eq.~\eqref{eq:Spherical harmonic trap}. Monopole coefficients $p_{n0}(n)$ versus principle number $n$ up to $n_{\text{max}}=14$.  Parameters: Trap frequency $\nu=\SI{22.1}{\hertz}$, particle number, chemical potential, aperture radius $R$: (a) $N=\num{10}$, $\mu/h =\SI{35.6}{\hertz}$, $R=\SI{9.21}{\micro \meter}$; (b) $N=\num{1000}$, $\mu / h=\SI{55.8}{\hertz}$, $R=\SI{8.95}{\micro \meter}$; (c) $N=\si{10^5}$, $\mu/h=\SI{289.2}{\hertz}$, $R=\SI{14.1}{\micro \meter}$. For the reconstruction of the density in Fig.~\ref{fig:radial_decomposed_dens_iso} we mark the cutoff $p_c=10^{-6}$ (gray \dashed).} 
    \label{fig:Monopole coefficients GP density isotropic}
\end{figure*}
\begin{figure*}
    \centering
    \includegraphics{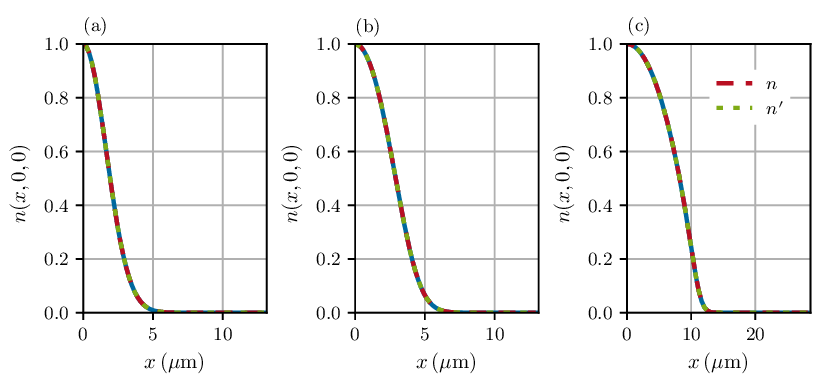}
    \caption{Cross-sections of the scaled Gross-Pitaevskii ground-state density distribution $n(x,0,0)$ versus Cartesian coordinate $x$ in a three-dimensional isotropic harmonic oscillator potential. Gross-Pitaevskii solution (blue \full). Interpolation of the density with Stringari polynomials $n(\br)$~Eq.~\eqref{eq:multipole expansion density} (red  \dashed) and alternatively with the cutoff $\mathfrak{p}_c=10^{-6}$  $n'(\br)$ (green \dotted). Parameters as in Fig.~\ref{fig:Monopole coefficients GP density isotropic}.}
    \label{fig:radial_decomposed_dens_iso}
\end{figure*}
We consider atomic density distributions in an isotropic harmonic oscillator potential~\eqref{eq:Spherical harmonic trap}. As the symmetry of the external potential determines the symmetry of the density, the condensate is interpolated by monopoles only. The efficiency of the interpolation depends on the actual radial shape $n(\br)=n(r)$, which will be determined by either the Thomas-Fermi density~\eqref{eq:Thomas-Fermi density} or the stationary Gross-Pitaevskii equation~\eqref{eq:stationary GP}. For the Gross-Pitaevskii density, we also evaluate the cumulant expansion to investigate the effect of different mean-field interactions.
\subsubsection{Thomas-Fermi density}
As the Thomas-Fermi density is directly proportional to the trapping potential, the interpolation is obtained by two Stringari polynomials only, as discussed in Sec.~\ref{sec:Harmonic, isotropic, three-dimensional oscillator}. Using the Thomas-Fermi approximation in its dimensionless form
\begin{equation}\label{eq:Thomas-Fermi parabola}
    n_\tf(\br)=\frac{15 N}{8 \pi} n'(r'), \quad  n'(r')=(1 - r'^2),
\end{equation}
with dimensionless radial coordinate $r'=r/r_\tf$ and the Thomas-Fermi radius $r_\tf=\sqrt{2\mu_\tf(N)/(M\omega^2)}$. The chemical potential scales with the particle number as
\begin{equation}
    \mu(N)= \frac{\hbar \omega}{2}\left(\frac{a_s N }{\ell}\right)^{2/5}, \quad \ell = \sqrt{\hbar/(M\omega)}.
\end{equation}
We find the monopole coefficients,
\begin{equation}\label{eq:Thomas-Fermi coefficients}
    n'_{000} = \frac{4}{5}\sqrt{\frac{\pi}{3}},\quad n'_{100} = \frac{4}{5}\sqrt{\frac{\pi}{7}}.
\end{equation}
\subsubsection{Gross-Pitaevskii density}
\label{subsec:Gross-Pitaevskii density}
\begin{figure*}
    \centering
    \includegraphics{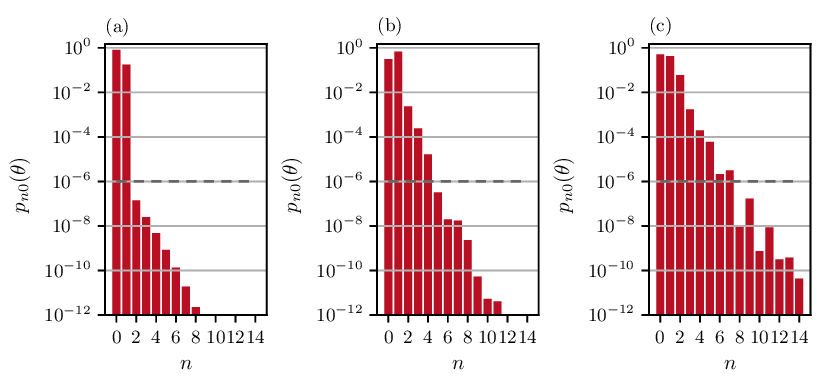}
    \caption{Multipole expansion of the isotropic Gross-Pitaevskii cumulant $\theta(\br)$ Eq.~\eqref{eq:multipole expansion cumulant density}. Monopole coefficients $p_{n0}(\theta)$ versus principle number $n$ with $n_{\text{max}}=14$. For the reconstruction of the cumulant in Fig.~\ref{fig:Cumulant of the isotropic ground-state density} we mark the cutoff $p_c=10^{-6}$ (gray \dashed). Parameters as in Fig.~\ref{fig:Monopole coefficients GP density isotropic}.}
    \label{fig:Monopoles of the cumulant of the isotropic Gross-Pitaevskii}
\end{figure*}
\begin{figure*}
    \centering
    \includegraphics{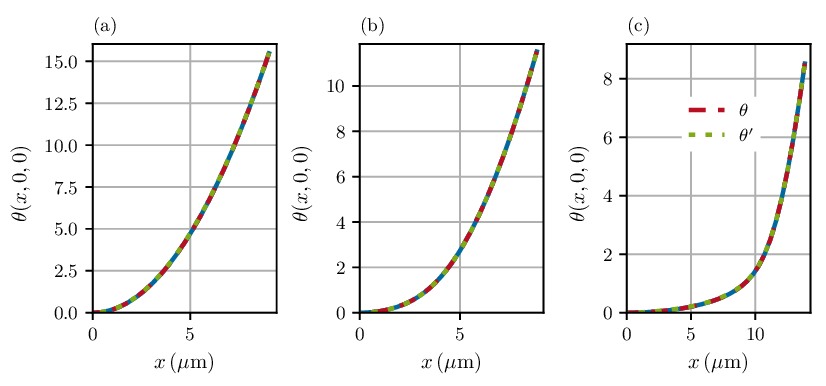}
    \caption{Cross-sections of the cumulant of the isotropic ground-state density distribution $\theta(x,0,0)$ versus Cartesian coordinate $x$ in a three-dimensional harmonic oscillator potential. Cumulant evaluated up to the aperture radius $R$. Gross-Pitaevskii solution (blue \full), interpolation of the cumulant with Stringari polynomials $\theta$ (red  \dashed) and alternatively with the cutoff $p_c=10^{-6}$ $\theta'$ (green \dotted). Parameters as in Fig.~\ref{fig:Monopole coefficients GP density isotropic}.}
    \label{fig:Cumulant of the isotropic ground-state density}
\end{figure*}
We represent the three-dimensional matter-wave field on a discrete, Cartesian Fourier grid. Therefore, we solve the stationary Gross-Pitaevskii equation~\eqref{eq:stationary GP} using Fourier spectral methods. To study the behavior of the radial expansion coefficients in different interaction regimes, we choose different particle numbers for the condensate. The $n_{nlm}$'s in the multipole expansion of density in Eq.~\eqref{eq:multipole expansion density} are obtained using the method of least-squares (see Eq.~\eqref{eq:least squares Stringaris}) replacing the target potential with the numerical, discrete Gross-Pitaevskii target density $n_t(\br_j)$. In contrast to the Thomas-Fermi density the radius $R=r_\tf$ of the spherical integration volume in Eq.~\eqref{eq:Stringaris} is not known a priori. Therefore, we are minimizing the least-square error $\epsilon(R)$ with respect to a variable aperture radius $R$
\begin{equation}
   \text{min}~ \varepsilon(R) = || \mat\Str(R) \vec{n} - \vec{n}_t||^2,
\end{equation}
for a fixed number of basis functions. The monopole coefficients $p_{n0}(n)$ for the isotropic Gross-Pitaevskii density are depicted in Fig.~\ref{fig:Monopole coefficients GP density isotropic} on a semi-logarithmic scale for $n_{\text{max}}=\num{14}$ basis functions. For small particle numbers \ref{fig:Monopole coefficients GP density isotropic} (a), the coefficients are declining exponentially for $n>6$ whereas the magnitude of the tenth coefficient is $p_{10,0}(n)<10^{-6}$. Entering the intermediate and strong interacting regime, the decline in magnitude becomes more irregular, as the Gaussian-like shape of the density is modified towards a polynomial shape. In Fig.~\ref{fig:Monopole coefficients GP density isotropic} (b), the most contributions to the polynomial expansion are within the first five coefficients, and in Fig.~\ref{fig:Monopole coefficients GP density isotropic} (c) within the first three. The latter reflects the transition to the pure quadratic Thomas-Fermi regime. However, we also recognize that the magnitude of the expansion coefficients does not converge to the same level as in the low interacting regime. In contrast to the Thomas-Fermi solution, the density also contains high-energetic modes close to the Thomas-Fermi radius, whose interpolation requires a lot of Stringari polynomials. As we are interested in good interpolation in the region of significant density, we introduce a cutoff at
\begin{equation}
    p_{\text{c}}=\frac{|n_{nlm}|^2}{P(n)} = 10^{-6},\quad  P(n)=\sum_{nlm}|n_{nlm}|^2,
\end{equation}
which disregards some of the highly energetic modes. In Fig.~\ref{fig:radial_decomposed_dens_iso}, we plot the corresponding expansion in terms of the polynomials which matches the Gross-Pitaevskii density quite well. We also use the Stringari polynomials with a reduced number of basis functions, neglecting coefficients smaller than the chosen cutoff. 

Besides the multipole coefficients for the density, we also look into the cumulant expansion in Eq.~\eqref{eq:multipole expansion cumulant}, for which we expect a faster convergence in the low-interacting limit. The $p_{n0}(\theta)$ in Fig.~\ref{fig:Monopoles of the cumulant of the isotropic Gross-Pitaevskii} (a) confirm that the density distribution is more of Gaussian shape, as the cumulant expansion almost terminates for monopole powers $n>3$. For larger particle numbers, subfigures (b) and (c), the cumulant expansion works quite efficiently as the polynomial series converges faster as in Fig.~\ref{fig:Monopole coefficients GP density isotropic}. Cross-sections of the cumulant and the interpolation with the Stringari polynomials are shown in Fig.~\ref{fig:Cumulant of the isotropic ground-state density}. 

\subsection{Anisotropic three-dimensional density}
\label{sec:3d anisotropic harmonic oscillator potential}
As discussed in Ref.~\cite{TESKE_2018_MeanfieldWigner}, the Thomas-Fermi field in a general harmonic oscillator potential can always be re-scaled to an isotropic s-wave by an affine coordinate transformation. Hence, this simplifies the search for the optimal aperture radius $R$ and adapts the polynomial expansion on the finite interval to the anisotropic extension of the density distribution. The latter becomes necessary for an optimal and efficient interpolation of the Gross-Pitaevskii matter-wave field which reaches beyond the Thomas-Fermi radius. 

For this purpose, we evaluate the covariance matrix 
\begin{align}
\mat{\Sigma}_{\br}=\braket{(\br-\br_0)\otimes (\br-\br_0)},\\ \vec{r}_0 = \braket{\br} = \int \rmd^3 r~  \br\,  n(\br),
\end{align}
for the three-dimensional Gross-Pitaevskii field. 
The positive, semi-definite matrix $\mat{\Sigma}_{\br}$ admits a Cholesky decomposition of the form
\begin{equation}
	\mat{\Sigma}_{\br} = \mathsf{C}\mathsf{C}^\top, \quad    \mathsf{C} = \mat{Q}\mat{\sigma}.
\end{equation}
The matrices $\mat{Q}$ and $\mat{\sigma}$ are defined by the eigenvalue equation
\begin{equation}
	    \mat{\Sigma} \mat{Q}= \mat{Q} \mat{\sigma}^2.
\end{equation}
The matrix $\mat{C}$ and the expectation value $\br_0$ define the required affine coordinate transformation
\begin{equation}\label{eq:Affine trafo Cholesky}
	\vec{\zeta}=\mathsf{C}^{-1} (\vec{r} - \vec{r}_0 ),
\end{equation}
that we use to evaluate a multipole expansion of the form
\begin{align}\label{eq:Multipole expansion scaled frame}
\begin{aligned}
	n(\br) = \sum_{n=0}^{\infty}\sum_{l=0}^\infty\sum_{m=-l}^l \mathfrak{n}_{nlm}S_{nlm}(\vec{\zeta}),\\
	\theta(\br) = \sum_{n=0}^{\infty}\sum_{l=0}^\infty\sum_{m=-l}^l \Uptheta_{nlm}S_{nlm}(\vec{\zeta})
	\end{aligned}
\end{align}
where the Stringari polynomials are evaluated with respect to the new coordinates $\vec{\zeta}$. For the new multipole coefficients $\mathfrak{n}_{nlm}$, $\Uptheta_{nlm}$ we define also the corresponding spectral powers
\begin{align}
    \mathfrak{p}_{nl}(n) &= \sum_{m=-1}^l \frac{|\mathfrak{n}_{nlm}|^2}{\mathfrak{P}(n)},\quad  \mathfrak{P}(n) = \sum_{nlm} |\mathfrak{n}_{nlm}|^2, \\
    \mathfrak{p}_{nl}(\theta) &= \sum_{m=-1}^l \frac{|\Uptheta_{nlm}|^2}{\mathfrak{P}(\theta)},\quad  \mathfrak{P}(\theta) = \sum_{nlm} |\Uptheta_{nlm}|^2,
\end{align}
\paragraph{Thomas-Fermi density}
\begin{figure}
\includegraphics[scale=1]{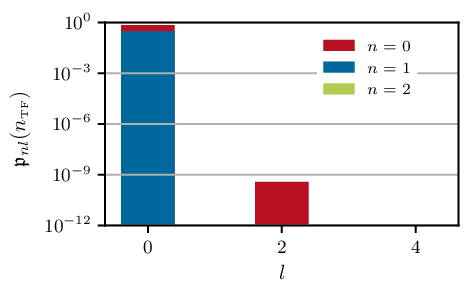} 
\caption{Multipole expansion of the scaled Thomas-Fermi density $n_\tf(\br)$ for the spheroidal harmonic oscillator Eq.~\eqref{eq:spheriodal harmonic trap} with anisotropy $\alpha=2$. Relative angular powers $\mathfrak{p}_{nl}(n_\tf)$ versus principle number $n$. Different principle numbers: red $n=0$, blue $n=1$, green $n=2$. $n_{\text{max}}=14$, $l_{\text{max}}=4$, $R=\SI{6.6}{\micro \meter}$.}
\label{fig:Relative angular powers for the Thomas-Fermi density in an anisotropic harmonic oscillator}
\end{figure}

\begin{figure}
\includegraphics[scale=1]{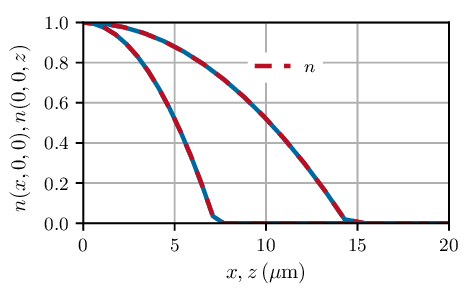} 
\caption{Cross-sections of the scaled Thomas-Fermi density $n_\tf(x,0,0)$, $n_\tf(0,0,z)$ (blue \full) versus Cartesian coordinates $x,\,z$ in a spheroidal harmonic oscillator Eq.~\eqref{eq:spheriodal harmonic trap}. Interpolation of the density with Stringari polynomials $n(\br)$ Eq.~\eqref{eq:multipole expansion density} (red  \dashed). Parameters: $\alpha = 2$, Thomas-Fermi radii $x_\tf=y_\tf=\SI{14.4}{\micro \meter}$, $z_\tf=\SI{7.2}{\micro \meter}$, particle number $N=\si{10^5}$, chemical potential $\mu_\tf / h=\SI{304}{\hertz}$.}
\label{fig:Cross-section of the exact interpolation of the Thomas-Fermi density cylindrical symmetry}
\end{figure}

\begin{figure*}
\centering
\includegraphics[scale=1]{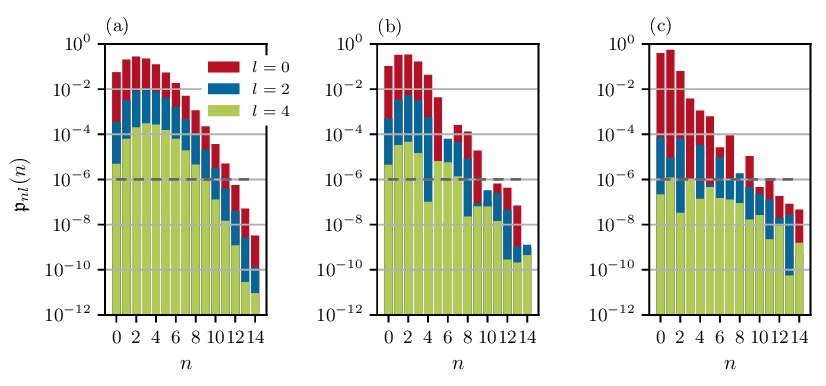}
\caption{Multipole expansion of the scaled Gross-Pitaevskii density $n(\br)$ Eq.~\eqref{eq:stationary GP} for the spheroidal harmonic oscillator Eq.~\eqref{eq:spheriodal harmonic trap} with anisotropy $\alpha=2$. Relative angular powers $\mathfrak{p}_{nl}(n)$ versus principle number $n$. Different angular momenta: red $l=0$, blue $l=2$, green $l=4$ with $n_{\text{max}}=14$, $l_{\text{max}}=4$. For the reconstruction of the density in Fig.~\ref{fig:Ground-state density distributions anisotropic harmonic oscillator} we mark the cutoff $\mathfrak{p}_c=10^{-6}$. Particle number, chemical potential, aperture radius: (a) $N=\num{10}$, $\mu/h =\SI{37.2}{\hertz}$, $R=\SI{11.6}{\micro \meter}$; (b) $N=\num{1000}$, $\mu / h=\SI{60.6}{\hertz}$, $R=\SI{12.1}{\micro \meter}$; (c) $N=\si{10^5}$, $\mu/h=\SI{307}{\hertz}$, $R=\SI{17.6}{\micro \meter}$.}
\label{fig:Different relative angular powers GP ansio}
\end{figure*}

\begin{figure*}
\centering
\includegraphics[scale=1]{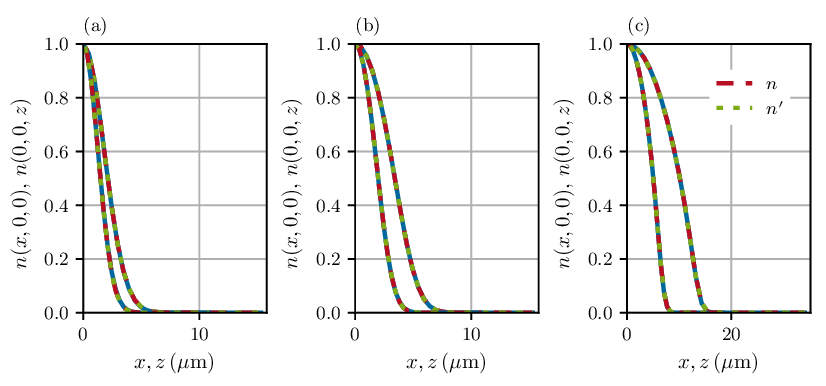}
\caption{Cross-sections of the scaled ground-state density distributions $n(x,0,0)$, $n(0,0,z)$ versus Cartesian coordinates $x,\,z$ in a spheroidal harmonic oscillator Eq.~\eqref{eq:spheriodal harmonic trap}. Gross-Pitaevskii solution (blue \full). Interpolation of the density with Stringari polynomials $n(\br)$~Eq.~\eqref{eq:multipole expansion density} (red  \dashed), alternatively with the cutoff $n'(\br)$ at $\mathfrak{p}_c=10^{-6}$ (green \dotted). Parameters as in Fig.~\ref{fig:Different relative angular powers GP ansio}.}
\label{fig:Ground-state density distributions anisotropic harmonic oscillator}
\end{figure*}

As a benchmark test, we investigate the Thomas-Fermi density in an anisotropic harmonic oscillator with cylindrical symmetry, which we discussed in Sec.~\ref{sec:Prolate or oblate spheriodal potential}. For the ratio of angular frequencies, we use $\alpha=2$. From analyzing the potential, we know that the multipole expansion in Eq.~\eqref{eq:multipole expansion cumulant density} just exhibits monopoles as well as one quadrupole. Applying the coordinate transformation Eq.~\eqref{eq:Affine trafo Cholesky}, we obtain the angular powers $\mathfrak{p}_{nl}(n_\tf)$ shown in Fig.~\ref{fig:Relative angular powers for the Thomas-Fermi density in an anisotropic harmonic oscillator}. As the multipole expansion is now performed in the scaled reference frame \eqref{eq:Multipole expansion scaled frame}, where the ellipsoid is re-scaled to a sphere, we expect monopoles only, see Eq.~\eqref{eq:Thomas-Fermi coefficients}. Indeed, we find good agreement with the isotropic Thomas-Fermi density as the quadrupoles are $\mathfrak{p}_{n2}(n_\tf)<10^{-9}$ as displayed in Fig.~\ref{fig:Relative angular powers for the Thomas-Fermi density in an anisotropic harmonic oscillator}. Using the monopoles $\mathfrak{n}_{n00}$ and the quadrupoles within the transformation matrix $\mat{C}$, one can reconstruct the original oblate-shaped Thomas-Fermi density as depicted in Fig.~\ref{fig:Cross-section of the exact interpolation of the Thomas-Fermi density cylindrical symmetry}.
\begin{figure*}
\centering
\includegraphics[scale=1]{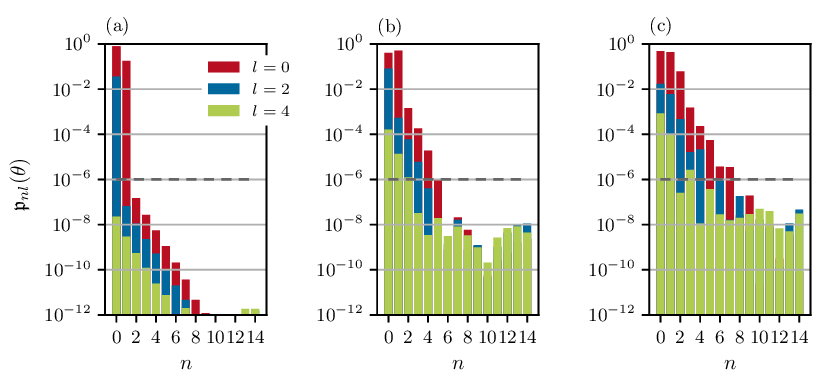}
\caption{Multipole expansion of the Gross-Pitaevskii cumulant $\theta (\br)$ Eq.~\eqref{eq:multipole expansion cumulant density} for the spheroidal harmonic oscillator Eq.~\eqref{eq:spheriodal harmonic trap} with anisotropy $\alpha=2$. Relative angular powers $\mathfrak{p}_{nl}(\theta)$ versus principle number $n$. Different angular momenta: red $l=0$, blue $l=2$, green $l=4$ with $n_{\text{max}}=14$, $l_{\text{max}}=4$. For the reconstruction of the cumulant in Fig.~\ref{fig:Cumulants ground-state density anisotropic} we mark the cutoff $\mathfrak{p}_c=10^{-6}$ (gray \dashed). Parameters as in Fig.~\ref{fig:Different relative angular powers GP ansio}.}
\label{fig:Different relative angular powers cumulant anisotropic} 
\end{figure*}
\begin{figure*}
\centering
\includegraphics[scale=1]{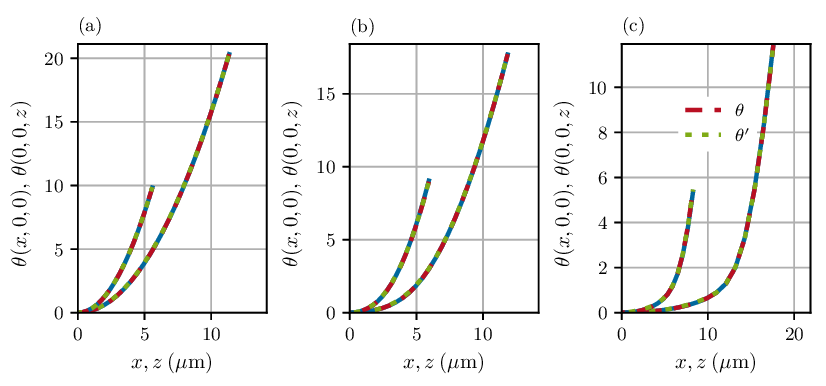} 
\caption{Cross-section of the cumulants $\theta(x,0,0)$,  $\theta(0,0,z)$ of the ground-state density distribution versus Cartesian coordinates $x,\,z$ in a spheroidal harmonic oscillator Eq.~\eqref{eq:spheriodal harmonic trap}. Cumulant evaluated up to the aperture radius $R$. Gross-Pitaevskii solution (blue \full). Interpolation of the cumulant with Stringari polynomials $\theta(\br)$ Eq.~\eqref{eq:multipole expansion cumulant} (red  \dashed), alternatively with the cutoff $\theta'(\br)$  at $\mathfrak{p}_c=10^{-6}$ (green \dotted). Parameters as in Fig.~\ref{fig:Different relative angular powers GP ansio}.
\label{fig:Cumulants ground-state density anisotropic}}
\end{figure*}
\paragraph{Gross-Pitaevskii density}
As in Sec.\ref{subsec:Gross-Pitaevskii density} we study the interpolation of the Gross-Pitaevskii density for different particle numbers varying the effective mean-field interaction in Eq.~\eqref{eq:stationary GP}. In addition, we compare the multipole expansion of the density with the multipole expansion of the cumulant. In both cases, the expansion coefficients are evaluated in the scaled reference frame defined by Eq.~\eqref{eq:Multipole expansion scaled frame}. In contrast to the ellipsoidal Thomas-Fermi density, we observe non-negligible quadrupole contributions in the relative angular powers $\mathfrak{p}_{nl}(n)$ for the low as well as for the high interacting regime, which is presented in Fig.~\ref{fig:Different relative angular powers GP ansio}. For low particle numbers, subfigure (a), the angular powers $\mathfrak{p}_{n0}(n)$, $\mathfrak{p}_{n2}(n)$, $\mathfrak{p}_{n4}(n)$ are decaying exponentially with respect to the principle number $n$ as we already stated in the isotropic case. Increasing the angular momentum for a  fixed value of $n$, the magnitudes of the $\mathfrak{p}_{nl}(n)$ decrease by roughly $1.5$-$2$ orders of magnitude. The angular momentum dependence decreases for increasing interactions as shown in the subfigures~\ref{fig:Different relative angular powers GP ansio} (b), (c). In particular, the powers $\mathfrak{p}_{n4}<10^{-6}$, are emphasizing the change of the Gross-Pitaevskii density towards the Thomas-Fermi shape. Moreover, the spectrum of the monopoles $\mathfrak{p}_{n0}(n)$ in the re-scaled reference frame exhibits the same structure as in the isotropic case (see Fig.~\ref{fig:Monopole coefficients GP density isotropic}), reflecting again the high-energetic modes in the Gross-Pitaevskii density that require a large number of Stringari polynomials.

Nevertheless, we can interpolate the ground-state density distributions also in the anisotropic harmonic oscillator as depicted in Fig.~\ref{fig:Ground-state density distributions anisotropic harmonic oscillator}. In particular, we can neglect modes with $l=4$ for the condensate with large particle numbers to obtain a good approximation with the Stringari polynomials.

The results for the anisotropic cumulant expansion are presented in Fig.~\ref{fig:Different relative angular powers cumulant anisotropic} and \ref{fig:Cumulants ground-state density anisotropic}. The monopoles $\mathfrak{p}_{n0}(\theta)$ exhibit again the same features as for the isotropic density, while the cumulant expansion works more efficiently describing the low-interacting regime. The latter is well described by just three multipole coefficients $\Uptheta_{000},\, \Uptheta_{020}$ and $\Uptheta_{200}$, subfigure~\ref{fig:Different relative angular powers cumulant anisotropic} (a). In contrast to the direct multipole expansion of the density, the cumulant expansion contains significant angular powers $\mathfrak{p}_{n4}(\theta)$ which needs to be considered for the polynomial interpolation.

\section{Release and free expansion of a Bose-Einstein condensate}
\label{sec:Expansion from magnetic chip traps}
Time of flight measurements is one of the standard techniques to image the density distribution of a Bose-Einstein condensate \cite{ketterle1999making} after a ballistic expansion and to extract equilibrium as well as dynamical properties. Here, we investigate the release of the condensate initially trapped in the Zeeman potential of the magnetic chip trap that we characterized in Sec.~\ref{sec:Magnetic Zeeman potential of an atom chip}.

Within the Thomas-Fermi approximation, Eq.~\eqref{eq:Thomas-Fermi density}, it is well-known \cite{CASTIN_1996_BoseEinsteinCondensates,KAGAN_1996_EvolutionBosecondensed,MEISTER_2017_EfficientDescription} that the time evolution of the density $n_\tf(\br,t)$ as well as the phase $\phi_\tf(\br,t)$ is given terms of the three adaptive scales $\lambda_j(t)$, $j=1,2,3$,
\begin{align}
n_\tf(\br,t) =& \frac{n_\tf(\{r_j/\lambda_j(t)\},0)}{\lambda_1 \lambda_2 \lambda_3},\label{eq:scaling TF density} \\
\phi_\tf(\br,t) =& \frac{M}{2 \hbar}\sum_{i=j}^3  r_j^2  \frac{\dot{\lambda}_j(t)}{\lambda_j(t)},
\label{eq:quadratic phase thomas-fermi}
\end{align}
if the condensate is initially trapped in a harmonic trap. The adaptive scales evolve during the ballistic expansion according to the differential equations
\begin{equation}
\label{eq:scales dgl free expansion}
\ddot{\lambda}_j = \frac{\omega_j^2(0)}{\lambda_j \lambda_1\lambda_2\lambda_3},
\end{equation}
with initial conditions chosen as $\lambda_j(0)=1$ and $\dot{\lambda}_j(0)=0$. Therefore, the conjugate variables $n_\tf(\br,t),\,\phi_\tf(\br,t)$ evolve quadratically in time making the multipole expansions with the introduced Stringari polynomials in Eq.~\eqref{eq:multipole expansion density},~\eqref{eq:multipole expansion phase} very efficiently, as only monopoles as well as quadrupole are contributing to Eq.~\eqref{eq:multipole expansion density} and Eq.~\eqref{eq:multipole expansion phase}. 
Beyond, the Thomas-Fermi approximation the initial Gross-Pitaevskii matter-wave field in the harmonic trap consists of non-quadratic polynomials
\begin{equation}
n(\br) = n_\tf(\br) + n_1(\br),
\end{equation}
as analyzed numerically by our multipole expansion in Sec.~\ref{sec:Multipole expansion of Bose-Einstein condensates} and described analytically in Ref.~\cite{FETTER_1998_ThomasFermiApproximation}. During the ballistic expansion of the condensate the density deviation $n_1(\br)$ leads to additional phase perturbations 
\begin{equation}\label{eq:total_phase}
\phi(\br,t) = \phi_\tf(\br,t)+ \phi_1(\br,t),
\end{equation}
to the quadratic Thomas-Fermi phase $\phi_\tf$ in Eq.~\eqref{eq:quadratic phase thomas-fermi}. 
We obtain the total phase in Eq.~\eqref{eq:total_phase} by solving the differential equation~\eqref{eq:scales dgl free expansion} for the adaptive scales $\lambda_j(t)$ and the Gross-Pitaevksii equation
\begin{align}
	\rmi \hbar \partial_t & \psi(\vec \xi,t)  = -\frac{\hbar^2}{2M}\sum_{j} \frac{1}{\lambda_j^2(t)}\partial^2_{\xi_j} \psi(\vec \xi,t)  \nonumber \\  +& \frac{1}{\lambda_1 \lambda_2 \lambda_3}\left(\frac{M}{2}\sum_j \omega^2_j(0)\xi_j^2 + g n(\vec{\xi},t) - \mu\right) \psi(\vec{\xi},t),
\end{align}
in the co-expanding frame of reference with the coordinates $\xi_j=r_j/\lambda_j(t)$. The transformed field $\psi(\vec{\xi},t)$ is related to the original one in Eq.~\eqref{eq:time-dependent GP equation} by 
\begin{align}
\Psi(\br,t) &= \frac{1}{\sqrt{\lambda_1\lambda_2\lambda_3}} \psi(\vec{\xi},t) \rme^{\rmi \left[ \phi_\tf(\vec \xi,t) - \beta(t)\right]},\\
\beta(t) &= \int^t \frac{\rmd t'}{\hbar} ~ \frac{\mu}{\lambda_1 \lambda_2 \lambda_3}.
\end{align}
At $t=0$, the field $\psi(\vec{\xi},0)$ satisfies the stationary Gross-Pitaevskii equation~\eqref{eq:stationary GP} with $N=\si{10^5}$ $^{87}$Rb atoms. As an external potential $U(\br)$ we choose the Zeeman potential $\UZeeman$ as well as the harmonic approximation. For the chosen parameter we find the Thomas-Fermi size $\br_\tf=(\num{25.5},\,\num{8.3},\,\num{9.4})\si{\micro \meter}$ and maximal density deviations of $\text{max}~\delta n=\num{0.1}\%$ within the two different trapping configurations.

\begin{figure}
\centering 
\includegraphics[scale=1]{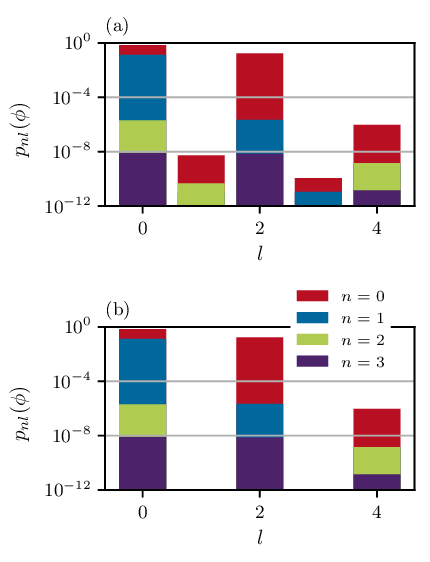}
\caption{Multipole expansion of the phase $\phi(\br,t_1)$ Eq.~\eqref{eq:total_phase} of an expanding Bose-Einstein condensate after $t_1=\SI{80}{\milli \second}$ time-of-flight. Relative angular powers $p_{nl}(\phi)$ Eq.~\eqref{eq:angular powers phi} versus angular momentum $l$. (a) condensate initially trapped in the Zeeman potential (see Sec.\ref{sec:Magnetic Zeeman potential of an atom chip}) of an atom chip, (b) condensate initially trapped in the anisotropic harmonic approximation the Zeeman potential Eq.~\eqref{eq:multipole components harmonic shifted trap}. Parameters of the trap as in Tab.~\ref{tab:release trap parameters}. Different principle numbers: red $n=0$, blue $n=1$, green $n=2$, purple $n=3$ with $n_{\text{max}}=3$, $l_{\text{max}}=4$.}
\label{fig:angular powers phase before lens}
\end{figure}

To analyze the impact on the phase during the ballistic expansion after a flight of flight $t_1=\SI{80}{\milli \second}$, we apply the multipole expansion of the total phase in Eq.~\eqref{eq:total_phase} with respect to the re-scaled coordinates $\vec{\xi}$ for these two different initial states. The relative angular powers $p_{nl}(\phi)$ 
\begin{align}
\label{eq:angular powers phi}
		p_{nl}(\phi) &=\sum_{m=-l}^l 
		\frac{|\phi_{nlm}|^2}{P(\phi)}
		,&
		P(\phi) &= \sum_{nlm} |\phi_{nlm}|^2,
\end{align}
are shown in Fig.~\ref{fig:angular powers phase before lens}. In the two subfigures, we compare the phase of the condensate that was initially trapped in the Zeeman potential (a) to the condensate initially in the harmonic approximation (b). From the latter, we state that the next leading orders to the scaling approximation \eqref{eq:quadratic phase thomas-fermi} are of the form $r^4$ with spectral powers $p_{20}(\phi)$, $p_{12}(\phi)$, and $p_{04}(\phi)$.  In addition, we find that their values are approximately three orders of magnitude higher than the anharmonic corrections which populate the dipole $p_{n1}(\phi)$, and the octupole moments $p_{n3}(\phi)$ in the phase of the condensate.
\section{Conclusion and outlook}
\label{sec:Conclusion and outlook}
In conclusion, we have introduced a multipole expansion with suitable radial polynomials to characterize different trapping geometries and the matter-wave field of a three-dimensional Bose-Einstein condensate. Besides the optical dipole potential for a single Laguerre-Gaussian beam, we have examined the multipole moments for the Zeeman potential of a realistic atom chip model. For both, we quantified deviations from their harmonic approximation and introduced an expansion of the cumulant which is superior for Gaussian-shaped functions. In the Thomas-Fermi approximation, the shape of the condensate is directly proportional to the external potential. Hence, it is natural to characterize the three-dimensional shapes of density and phase in terms of the same polynomial basis functions. Moreover, we have examined the efficiency of our multipole expansion for the different mean-field interactions in the Gross-Pitaevskii equation. In addition, we studied the phase of an expanding condensate in the same manner. We identified possible aberrations for long-time atom interferometry in the different multipole moments that are caused either by the external potentials or the intrinsic properties of interacting Bose-Einstein condensates. 

Our work provides a general and universal framework for an aberration analysis in matter-wave optics with interacting Bose-Einstein condensates. The multipole analysis allows the design for aberrations balanced matter-wave lenses in single or multiple lens setups \cite{DEPPNER_2021_CollectiveModeEnhanced,KANTHAK_2021_TimedomainOptics}, e.g. with programmable optical dipole potentials using digital micromirror devices \cite{Gauthier16}. Finally, the multipole expansion of the magnetic field could be used to exploit different trapping geometries and for designing new atom chips \cite{SACKETT_2023_TimeorbitingpotentialChip}.
\section*{Author Declarations}
\subsection*{Conflict of interest}
The authors have no conflict of interest to disclose. 
\section*{Data Availability}
The data that support the findings of this study are available from the corresponding author upon reasonable request.
\section*{Acknowledgements}
This work is supported by the DLR German Aerospace Center with funds provided by the Federal Ministry for Economic Affairs and Energy (BMWi) under Grant No. 50WM1957 and No. 50WM2250E. We acknowledge the members of the QUANTUS collaboration for continuous feedback. 
We thank A. Neumann, J. Battenberg, B. Zapf for their contributions to the python simulation package \textit{Matter Wave Sim} (MWS) implementing (3+1)~dimensional Bragg beam splitters with Gaussian laser beams \cite{NEUMANN_2021_AberrationsDimensional} and magnetic chip traps. 
\appendix   
\section{Jacobi polynomials}
\label{sec:Jacobi polynomials}
The Jacobi polynomials are defined by a Gaussian hypergeometric function ${}_{2}F_1(a,\alpha,\beta)$ \cite{OLVER_2010_NISTHandbook} for integer values $a=-n$,
\begin{align}\label{eq:Jacobi Hyper}
J^{(\alpha,\beta)}_{n}\left(x\right)=&\frac{{\left(\alpha+1\right)_{n}}}{n!} \times \nonumber \\
&{}_{2}F_1\left({-n,n+\alpha+\beta+1;\alpha+1};\frac{1-x}{2}\right),
\end{align}
where $(\cdot)_n$ denotes the Pochhammer symbol. They are orthogonal on the interval $x \in [-1, 1]$,
\begin{align}
    \int_1^1 & \rmd x ~ w_{\alpha,\beta}(x) J_n^{(\alpha, \beta)}(x) J_m^{(\alpha, \beta)}(x) =  \mathcal{A}_n \delta_{n,m},\label{eq:Jacobi orthogonality} \\
    \mathcal{A}_n&=\frac{2^{\alpha+\beta+1}}{2n+\alpha+\beta+1} \frac{\Gamma(n+\alpha+1)\Gamma(n+\beta+1)}{\Gamma(n+\alpha+\beta+1)n!},\label{eq:Jacobi normalization constant}
\end{align}
with respect to the weight function
\begin{equation}
	w_{\alpha,\beta}(x) = (1-x)^\alpha (1+x)^\beta.
\end{equation}
The Stringari polynomials in Eq.~\eqref{eq:Stringaris} are shifted Jacobi polynomials with $\alpha=l+1/2$ and $\beta=0$ substituting the coordinate as $x=1-2(r/R)^2$, $r \in [0, R]$. The normalization constant $\mathcal{N}_{nl}$ in Eq.~\eqref{eq:Stringaris} is obtained by using Eq.~\eqref{eq:Jacobi normalization constant}.
\section{Magnetic trapping on an atom chip}
\label{sec:Chip model}
\begin{figure}
\centering
\includegraphics[scale=0.4]{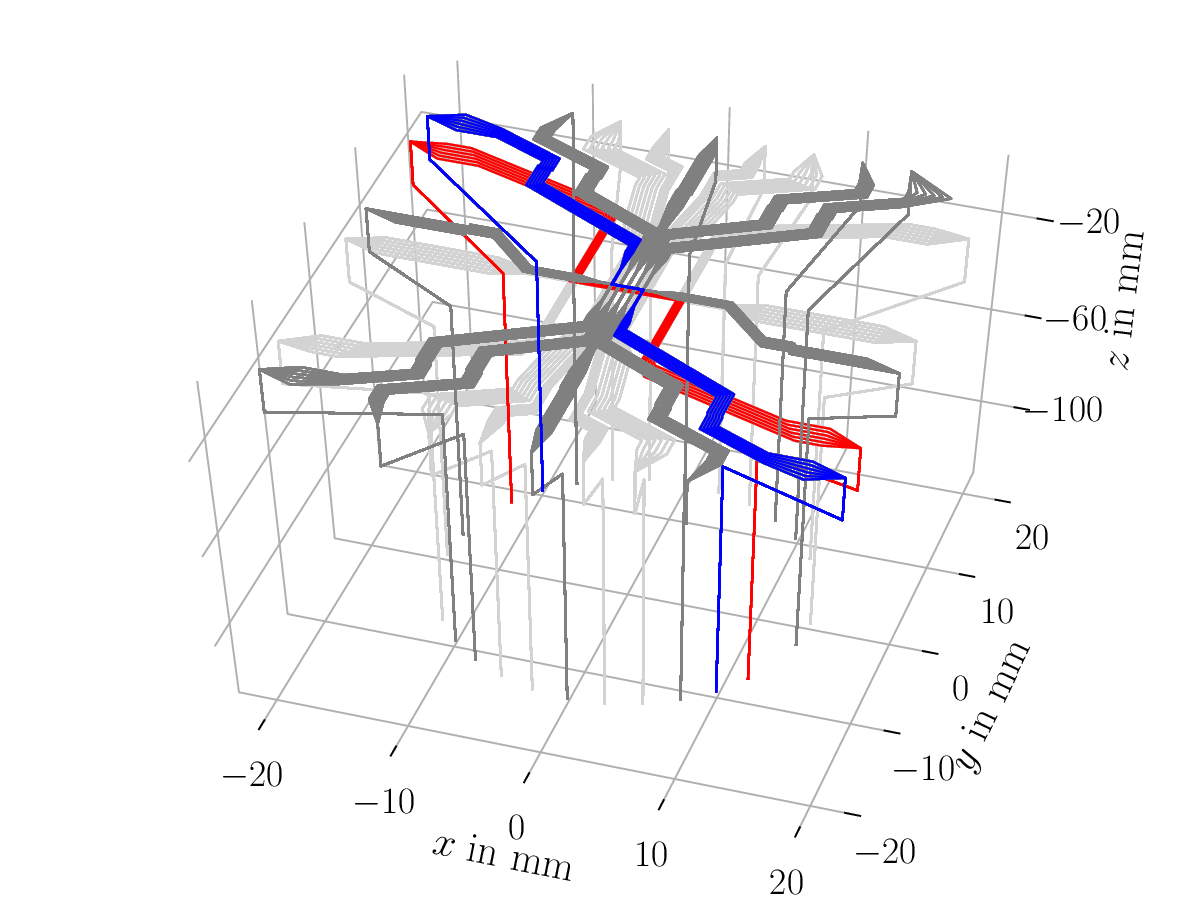} 
\caption{QUANTUS II atom chip model \cite{HERR_2013_KompakteQuelle, RUDOLPH_2015_HighfluxBEC}. Light-gray wires belong to the base chip structure. Gray wires belong to the science chip structure. Active conductors in Z-trap configuration in blue (science chip) and red (base chip) colors. External Helmholtz coils creating a homogeneous field are not depicted.}\label{fig:Q2-chip}
\end{figure}

The atom chip model is a representation of the experiment \cite{RUDOLPH_2015_HighfluxBEC} and is shown in Fig.~\ref{fig:Q2-chip}. The chip consists of three isolated conducting layers providing several possible trapping configurations. 
The first layer holds the largest mesoscopic structures. The U-shaped wires form a quadrupole field that is used for the three-dimensional magneto-optical trap (MOT). The second layer, the base chip (BC), and the third layer, the science chip (SC), consist of 4 and 5-wire two-dimensional strips, respectively, which intersect with one central orthogonal wire.
We regard the active conductors on the base as well as on the science chip in Z-trap configuration which are marked in red and blue colors in Fig.~\ref{fig:Q2-chip}. Both create an Ioffe-Pritchard-type trapping potential, that is used for releasing  and collimating the condensate. The field is superposed by a magnetic bias field $\vec{B}_0$ created by three pairs of Helmholtz coils.

\begin{figure}
\centering
\includegraphics[width=0.9\columnwidth]{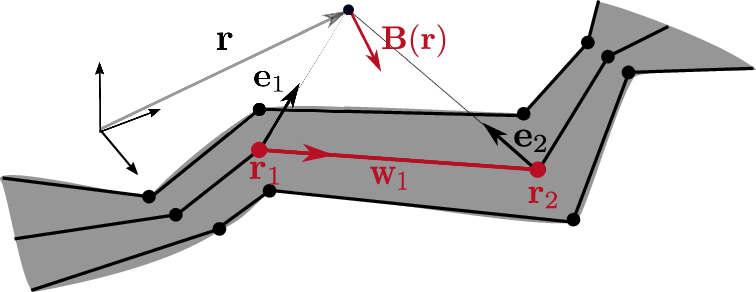}
\caption{Subsection with $N=5$ segments from a finite two-dimensional conducting strip of the QUANTUS II atom chip in Fig.~\ref{fig:Q2-chip}. Each segment is modeled by $M$ finite wires, here $M=3$.  The magnetic induction $\vec{B}(\br)$ at an observation point $\vec{r}$ created by a current $I_i$ in the finite wire element pointing into direction $\vec{w}_1$.}\label{fig:Biot-Savart}
\end{figure}

The magnetic induction field $\vec{B}_c$ of the atom chip is calculated by splitting $N$ two-dimensional wire strip segments into $M$ finite wire elements (cf. Fig.~\ref{fig:Biot-Savart}), that describe the shape of all wire strips in all layers. We use finite wires with lengths $l_i=|\vec{r}^{(i)}_{2}-\vec{r}^{(i)}_{1}|$ that point into the directions $\vec{w}^{(i)}=(\vec{r}^{(i)}_{2}-\vec{r}^{(i)}_{1})/l_i$ and carry a steady current $I_i=I / M$, where is $M$ is the number of wires representing a segment $i$  (see Fig.~\ref{fig:Biot-Savart}). The magnetic induction for a single finite wire element follows from Biot-Savart law\cite{JACKSON_2003_ElectrodynamicsClassical} using the parametrization of one wire $\br'(s)=\br_1 + \vec{w} l s$, $s \in [0,1]$, 
\begin{equation}
    \vec{B}_c(\br) = \frac{\mu_0}{4 \pi} \frac{I}{\varrho} 
\frac{\vec{e}_1 \times \vec{w}}{1-( \vec{e}_1\cdot \vec{w})^2}
\left(
\vec{e}_2
- \vec{e}_1 
\right)\cdot \vec{w},
\end{equation}
where the unit vectors $\vec{e}_{2},\,(\vec{e}_{1})$ are pointing from the conductor's end (start) to the observation point $\br$ and $\varrho=|\vec{r}-\vec{r}_{1}|$ denotes the distance to the head of the wire element at $\vec{r}_{1}$. Hence, the total magnetic induction field of $M$ wires in  $N$ segments is just the sum of all individual fields
\begin{align}
\begin{aligned}
\vec{B}(\br) &=  \vec{B}_0 +\sum_{i=1}^{NM} \vec{B}_c^{(i)}(\br)\\
\vec{B}_c^{(i)}(\br)&=\frac{\mu_0}{4 \pi}  \frac{I_i}{\varrho_i} 
\frac{\vec{e}^{(i)}_{1} \times \vec{w}_{}^{(i)}}{1-( \vec{e}^{(i)}_{1}\cdot \vec{w}^{(i)})^2}
\left(
\vec{e}^{(i)}_{2}
- \vec{e}^{(i)}_{1} 
\right)\cdot \vec{w}^{(i)}.
\end{aligned}
\label{eq:biot-savart-superposition}
\end{align}
\begin{table}
  \begin{tabular}{ll}
Wire & Current \\
    \toprule
Science chip &\SI{2.0}{\ampere} \\
Base chip & \SI{6.0}{\ampere} \\
x-coils  &\SI{0.1}{\ampere} \\
y-coils & \SI{-0.37431}{\ampere}\\
z-coils & \SI{0.0}{\ampere}\\
  \end{tabular}
  \caption{Typical set of currents applied to the atom chip to generate a magnetic trap.}
  \label{tab:release trap currents}
\end{table}
\begin{table}
  \begin{tabular}{lll}
Parameter & Symbol & Value \\
    \toprule
Spring constant  & $\vec{k}$ &  (\num{709}, \num{6685}, \num{5210}) \si{\kilo \hertz \per \square \milli \meter}\\
Frequencies $^{87}$Rb & $\vec{\nu}_\rb$ & (\num{9.08}, \num{27.88}, \num{24.61}) \si{\hertz}\\
Frequencies $^{41}$K & $\vec{\nu}_\ka$ & (\num{13.23}, \num{40.41}, \num{35.86}) \si{\hertz} \\
Trap minimum &$\br_0$ & (\num{0}, \num{0}, \num{1462}) \si{\micro \meter} \\
Tait-Bryan angles (XYZ) & $\alpha,\beta,\gamma$ & (\num{0.}, \num{0}, \num{9.7})\si{\degree} \\
  \end{tabular}
  \caption{Physical parameters of the QUANTUS II release trap. Spring constants and trap frequencies are corresponding to the magnetic substate $\ket{F=2,m_F=2}$. Tilt angles are evaluated in the chip coordinate system.
  }
  \label{tab:release trap parameters}
\end{table}
\section{Numerical evaluation}
\label{sec:Numerical evaluation}
While the multipole coefficients may be evaluated analytically using the scalar product in Eq.~\eqref{eq:Multipole coefficients by projection}, we are using a least-square evaluation \cite{NOCEDAL_2006_NumericalOptimization} when the potential is represented on a numerical grid. As the discretized Stringari polynomials are non-orthogonal basis functions, we introduce the finite complex scalar product
\begin{equation}
    (a|b) = \sum_{\vec{r}_j \in V} \mathcal{V}\, a^*(\br_j)b(\br_j)=\vec{a}^\dag \vec b,
\end{equation}
with the discrete position coordinates $\{\br_j\}$, the measure of the Cartesian volume element $\mathcal{V}= \Delta x \Delta y \Delta z$ and its norm $||\vec{a}|| = \sqrt{(a|a)}$. Hence, the distance of the squared residuals is given by
\begin{equation}\label{eq:least squares Stringaris}
    \varepsilon = || \mat\Str \vec{u} - \vec{U}||^2.
\end{equation}
Here, we have introduced the complex coefficient vector $\vec{u}=\{U_{nlm}\}$, the values of the discrete target potential $\vec{U}=\{U(\br_j)\}$ and the complex matrix $\mat{S}=\{S_{nlm}(\br_j)\}$ that contains the discrete set of the finite Stringari basis functions. One finds the least square minimum 
\begin{equation}
	\partial \varepsilon / \partial \vec u^* = 0,
\end{equation}
which leads to the best potential parameter estimate
\begin{equation}
    \bar{\vec u}  = (\mat \Str ^\dag \mat{\Str})^{-1}\mat \Str^{\dag} \vec{U}.
\end{equation}

\bibliography{bibliography}

\end{document}